\documentclass[%
reprint, 
 amsmath,
 amssymb,
 aps,
 prl,
 jor,
 sor,
 pra,
]{revtex4-2} 
\usepackage{amsmath}
\usepackage{amsthm}
\usepackage{ragged2e} 
\usepackage{subfig}
\usepackage{hyperref}
\usepackage{algorithm}
\usepackage{adjustbox}
\usepackage{algpseudocode}
\usepackage{etoolbox}
\usepackage{graphicx}
\usepackage{dcolumn}
\usepackage{bm}
\usepackage{float}
\usepackage{caption}
\usepackage{subcaption}     
\usepackage{physics}
\usepackage{mathtools}
\usepackage{hyperref}
\hypersetup{ 
     colorlinks=true, 
     linkcolor=blue, 
     filecolor=blue, 
     citecolor = blue,       
     urlcolor=blue, 
     } 

\usepackage[mathlines]{lineno}
\begin{document}
\title{Spectral Coarse-Graining and Rescaling for Preserving Structural and Dynamical Properties in Graphs}
\author{M. Schmidt}
\affiliation{ 
Department of Computer Science, University College London, UK}
\author{F. Caccioli}
\affiliation{ 
Department of Computer Science, University College London, UK}
\author{T. Aste}\affiliation{ 
Department of Computer Science, University College London, UK}
\begin{abstract}
We introduce a graph renormalization procedure based on the coarse-grained Laplacian, which generates reduced-complexity representations for characteristic scales identified through the spectral gap. This method retains both diffusion probabilities and large-scale topological structures, while reducing redundant information, facilitating the analysis of large graphs by decreasing the number of vertices. Applied to graphs derived from EEG recordings of human brain activity, our approach reveals macroscopic properties emerging from neuronal interactions, such as collective behavior in the form of coordinated neuronal activity. Additionally, it shows dynamic reorganization of brain activity across scales, with more generalized patterns during rest and more specialized and scale-invariant activity in the occipital lobe during attention-focused tasks.
\end{abstract}                              
\maketitle
\begin{figure}[H]
    \centering 
            \includegraphics[width=0.6\linewidth]{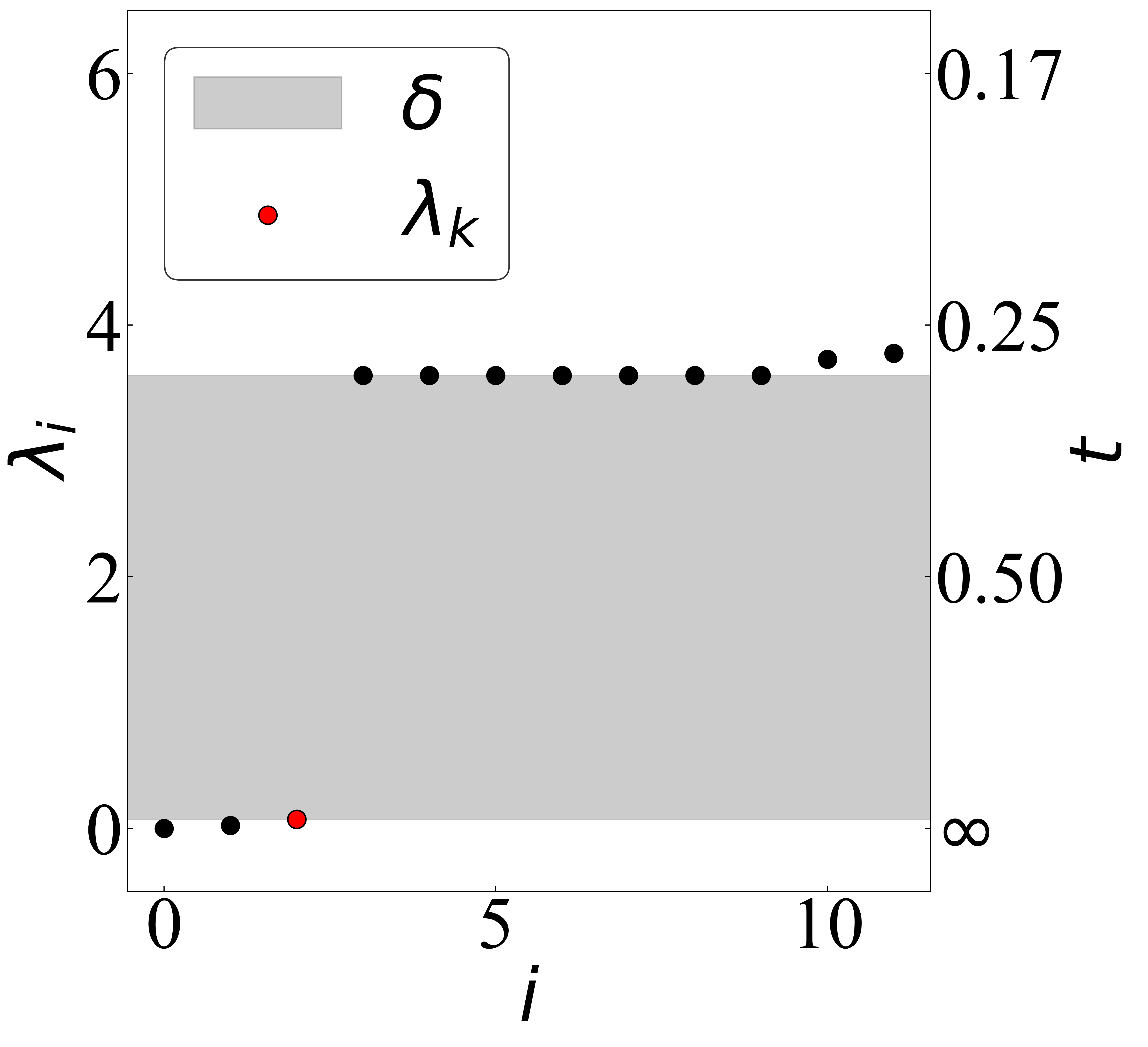} 
            \\
        \includegraphics[width=0.8\linewidth]{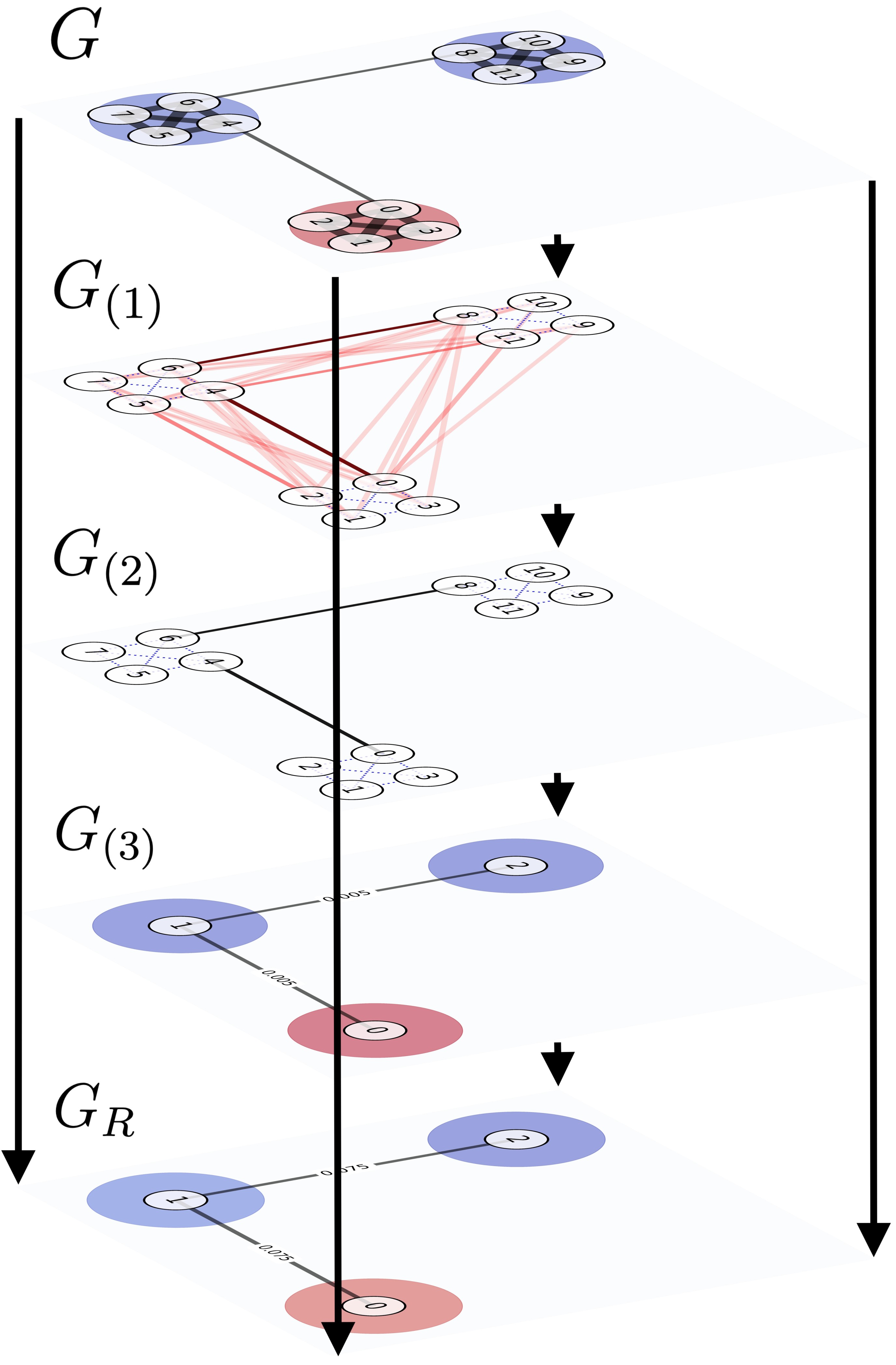} 
        \captionsetup{justification=justified, singlelinecheck=false, font=small}
        \caption{Spectral gap and renormalization of graph (a). 
        First, the spectral gap $\delta$ of the graph Laplacian of $G$ is identified, indicating the scale $\lambda_{k}$ to zoom out to using  Eq.~\ref{NewLaplacian}, which results in $G_{(1)}$. The matrix $G_{(1)}$ represents vertex similarities from a zoomed-out perspective, where positive weights (red line) indicate dissimilar vertices not connected in the original graph, while positive weights (black line) indicate dissimilar vertices connected in the original graph. Negative weights (blue dotted line) indicate similar vertices connected in the original graph. Thus, they can be contracted into effective vertices. This is achieved by assigning $G_{(1)}$'s weights to $G$ using Eq.~\ref{A2} resulting in $G_{2}$, and then contracting vertices connected by a blue edge into an effective vertex, resulting in $G_{3}$. The new effective weights reflect that fast diffusion modes have been removed by Eq.~\ref{NewLaplacian}. Thus, $G_{1}$ indicates how to derive a coarse-grained representation of the structure and dynamics of $G$. However, zooming out reduces resolution. To restore the original sharpness, we adjust the focus by rescaling the weights using Eq.~\ref{FinalKSRG_eq}, resulting in the renormalized graph $G_{R}$. The rescaling approximately preserves the original diffusion dynamics (see Fig.~\ref{DiffDySynNetworks}). 
        }
    \label{SynNetworks}
\end{figure}
\begin{figure}[H]
        \centering
        \begin{tabular}{c}
             Graph (a)\\
            \includegraphics[width=0.6\linewidth]{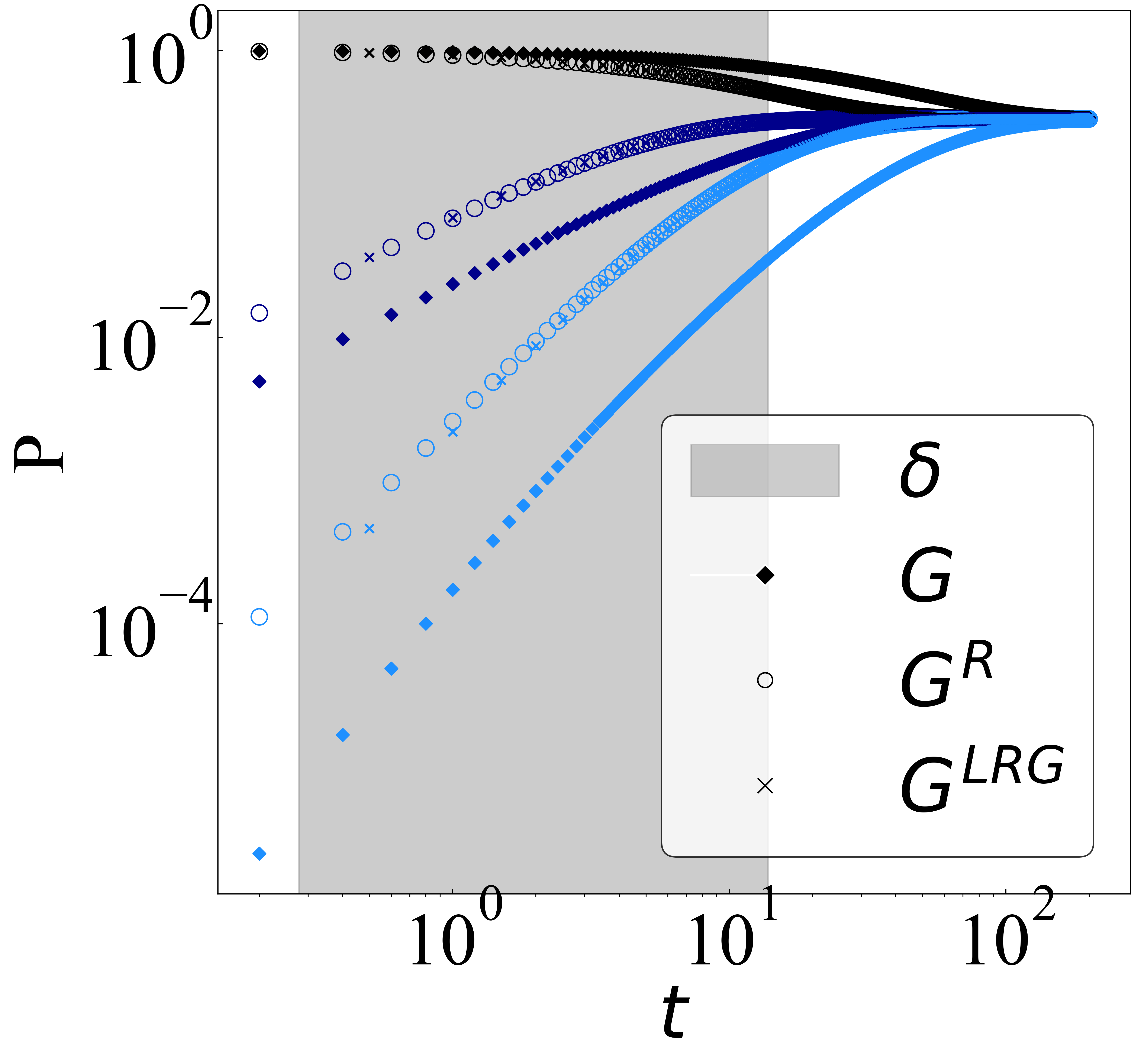}\\
             Graph (b) \\
            \includegraphics[width=0.6\linewidth]{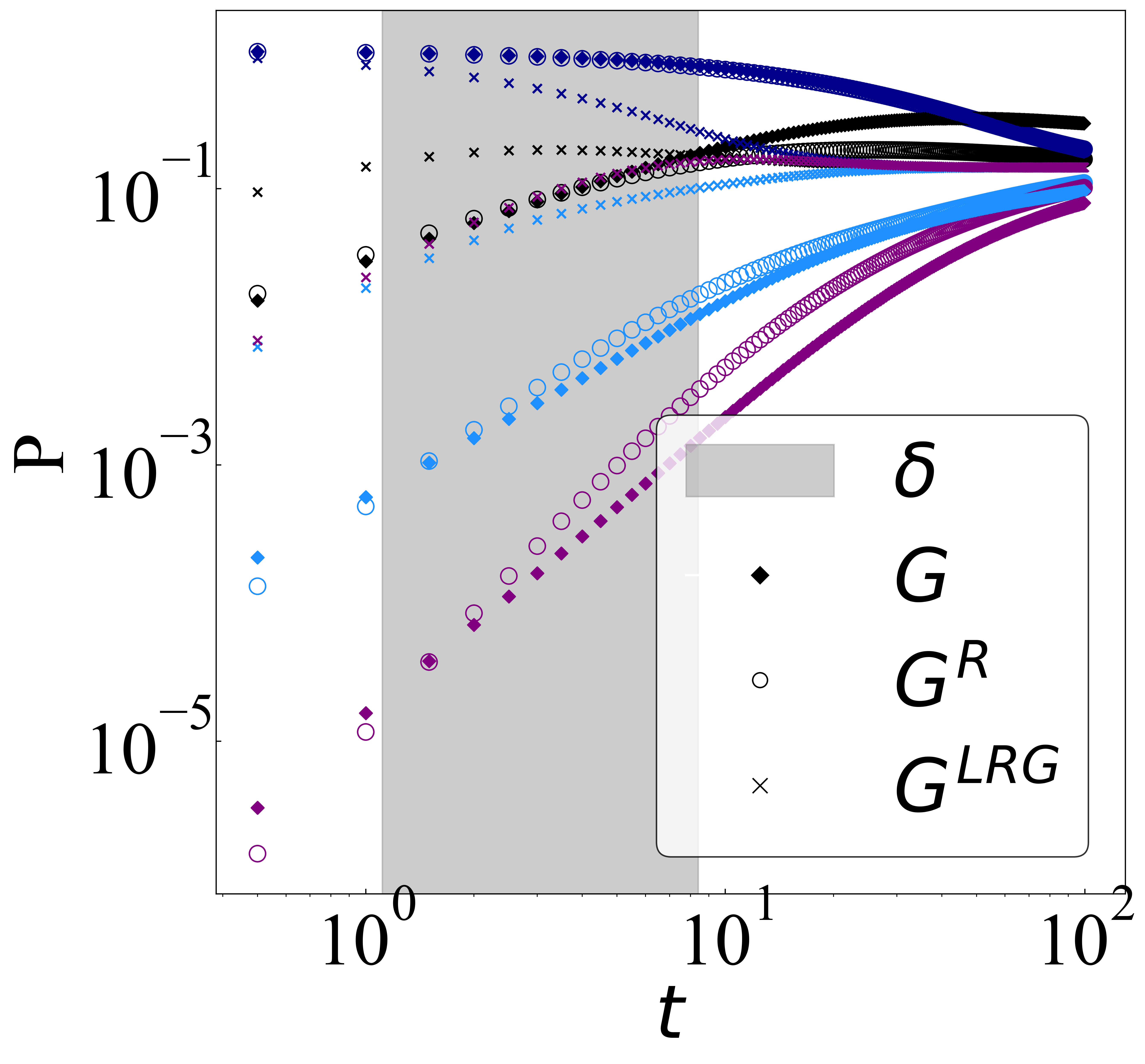}
        \end{tabular}
    \captionsetup{justification=justified, singlelinecheck=false, font=small}
    \caption{Diffusion dynamics. The probabilities $\mathbf{p}(t)$ of the clusters (original graph $G$) and effective vertices (renormalized graphs $G^{R}$ or $G^{LRG}$) are compared. These probabilities are determined by equation Eq.~\eqref{SolutionMaster}, with $\mathbf{p}(0)$ highlighted in red in Fig. \ref{SynNetworks} and Fig. \ref{BA_Net}. The graph $G^{R}$ is obtained using our renormalization procedure with $\lambda_{k}= 0.07$ for graph (a) and $\lambda_{k}=0.12$ for graph (b), while $G^{LRG}$ is obtained using \cite{Villegas_2023}'s Wilsonian renormalization procedure with $\lambda^{*} = 1/0.79$ for graph (a) and $\lambda^{*} = 1/2.79$ for graph (b) (identified by specific heat peak \cite{Villegas_2023}). Each color represents a different effective vertex or cluster. For (b), we consider the renormalized graph associated with the first spectral gap and omit the clusters and the corresponding effective vertices 3,4, and 6 for clarity. Closer alignment between points of the same color of $G^{R}$ or $G^{LRG}$ with $G$ indicates that the original diffusion dynamics are better preserved. The goodness with which our method preserves the diffusion dynamics depends on $\lambda_{k+1}$ and the spectral gap $\delta$ (see Eq.~\ref{Error}).}
    \label{DiffDySynNetworks}
\end{figure}
\begin{figure}[H]
    \centering 
        \includegraphics[width=0.6\linewidth]{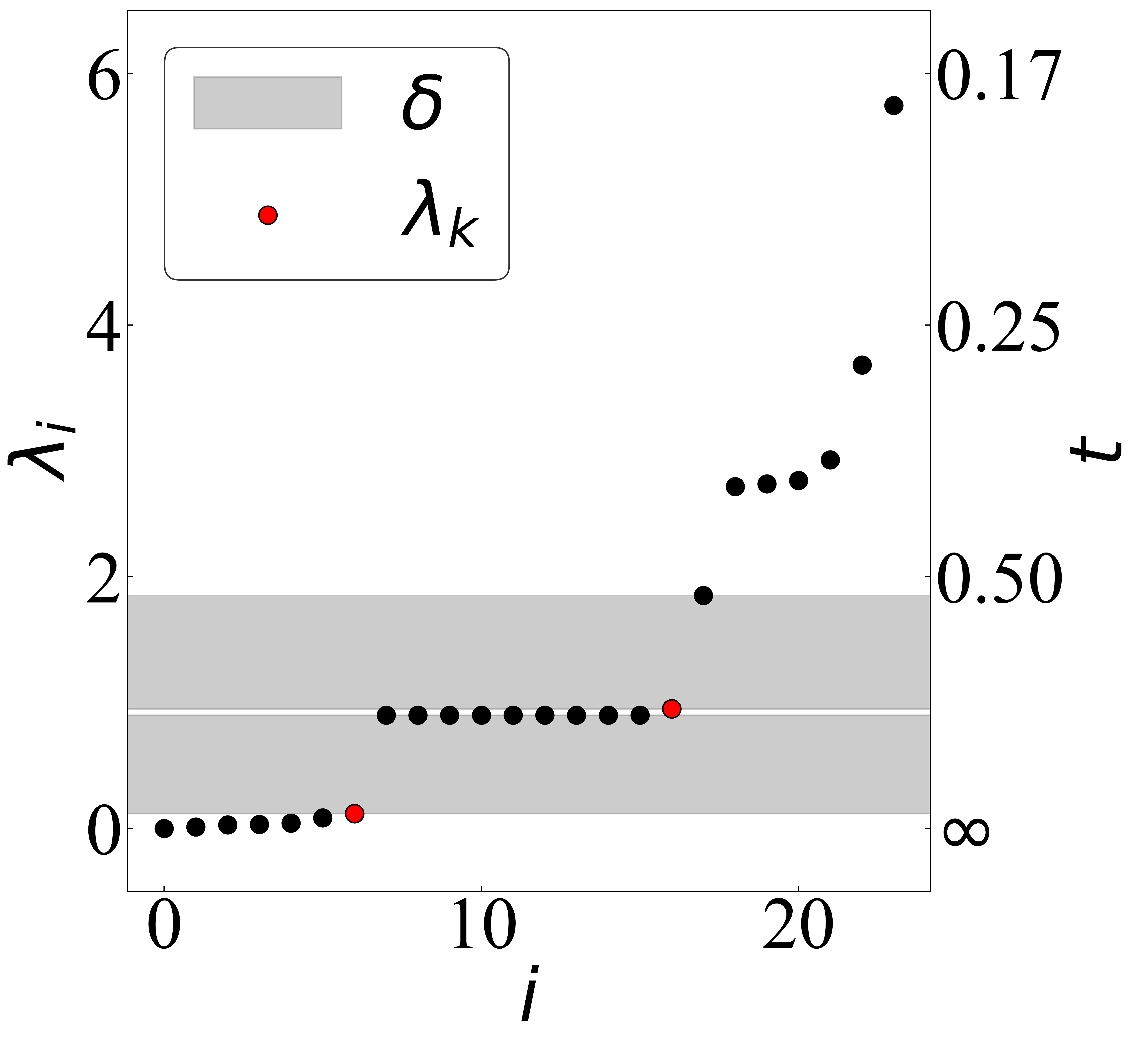} \\ 
        \includegraphics[width=0.8\linewidth]{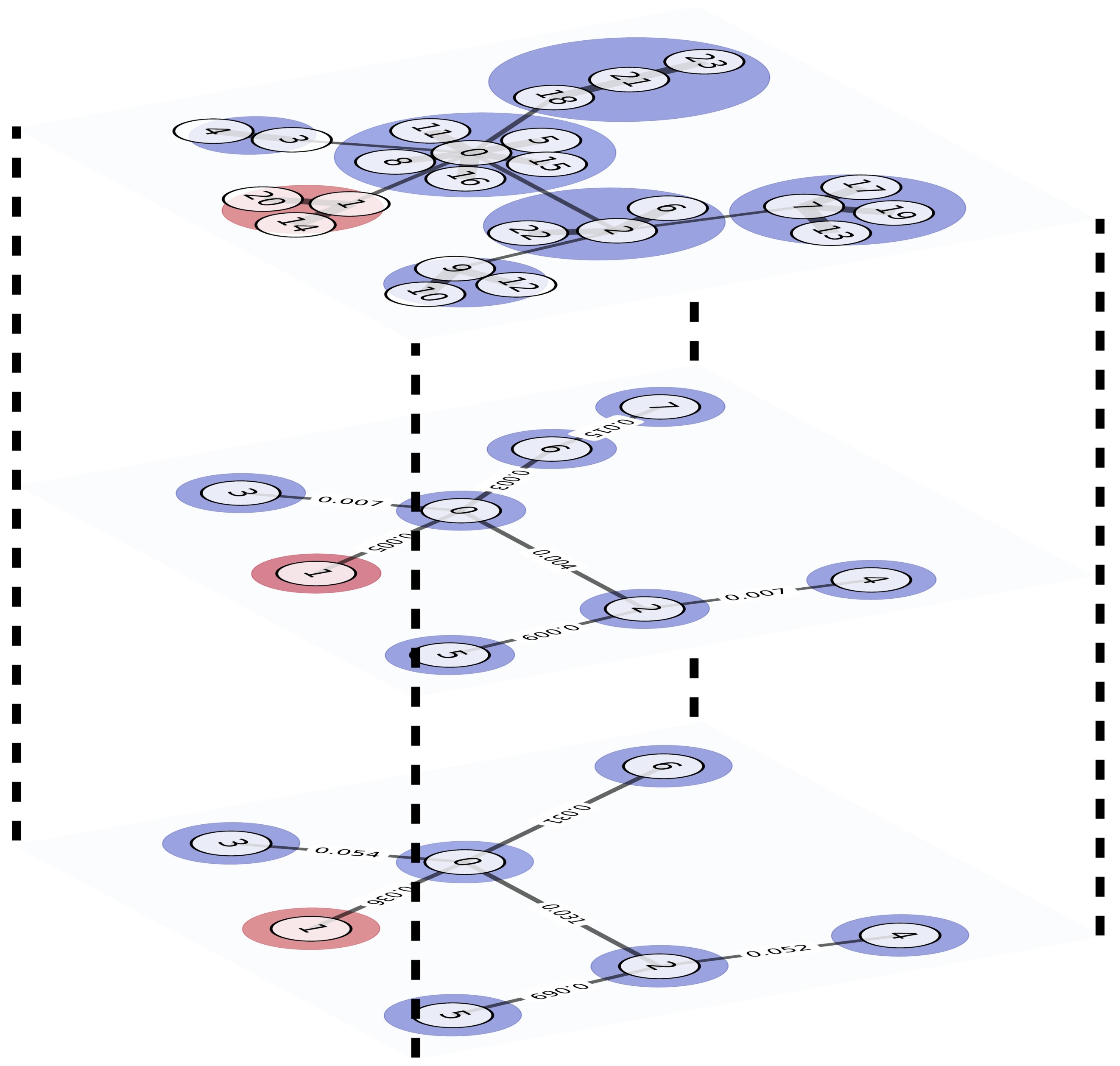} 
        \captionsetup{justification=justified, singlelinecheck=false, font=small}
        \caption{Spectral gap and renormalization of graph (b). A renormalized graph is obtained for each of the two spectral gaps. \cite{WILSON197475}.
        }
    \label{BA_Net}
\end{figure}
\begin{figure*}
    \begin{tabular}{cc}
             Rest &\hspace{3.5cm}Attention \\
            \includegraphics[width=0.3\linewidth]{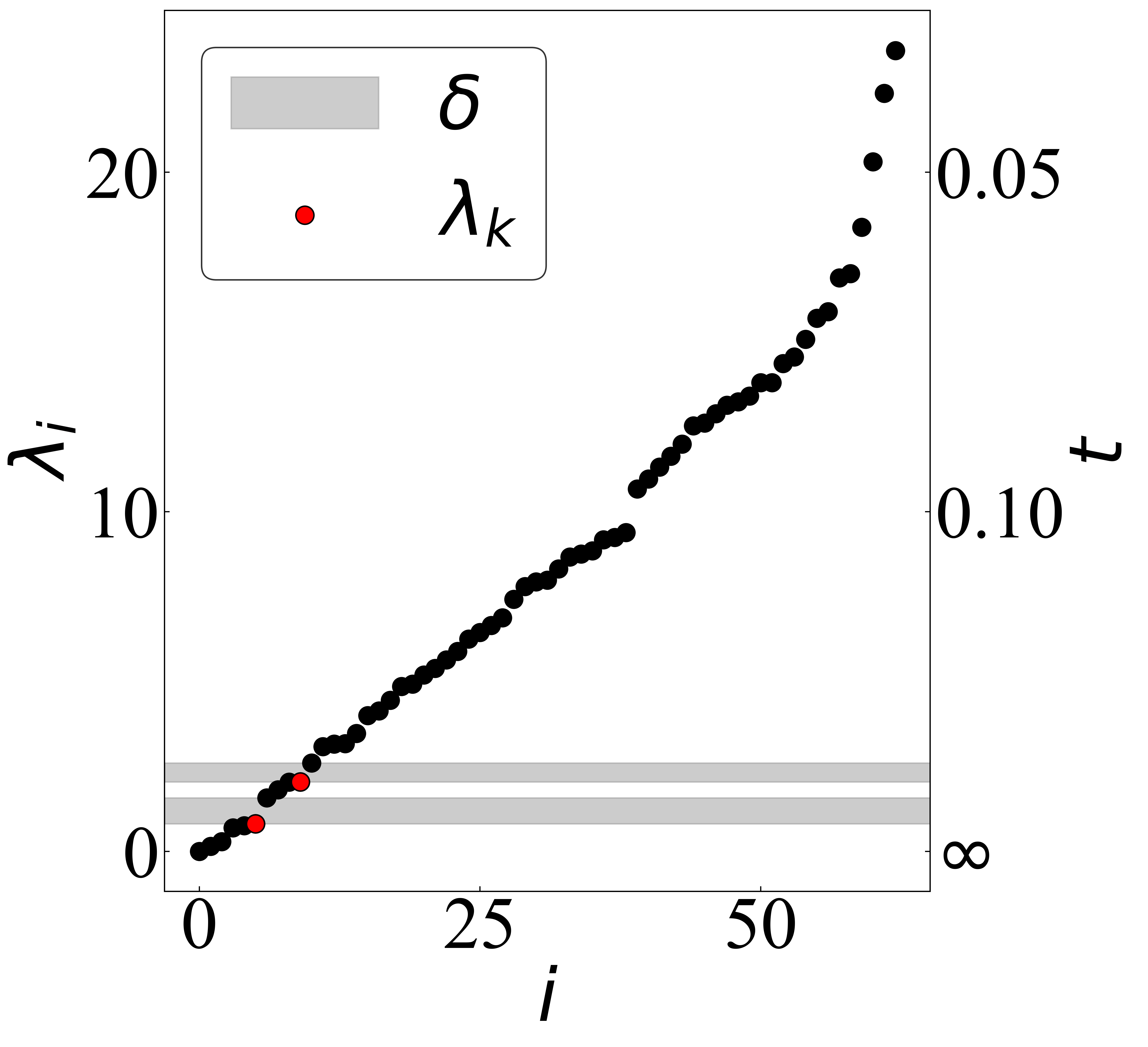}\hspace{3.5cm}&\hspace{3.5cm}
            \includegraphics[width=0.3\linewidth]{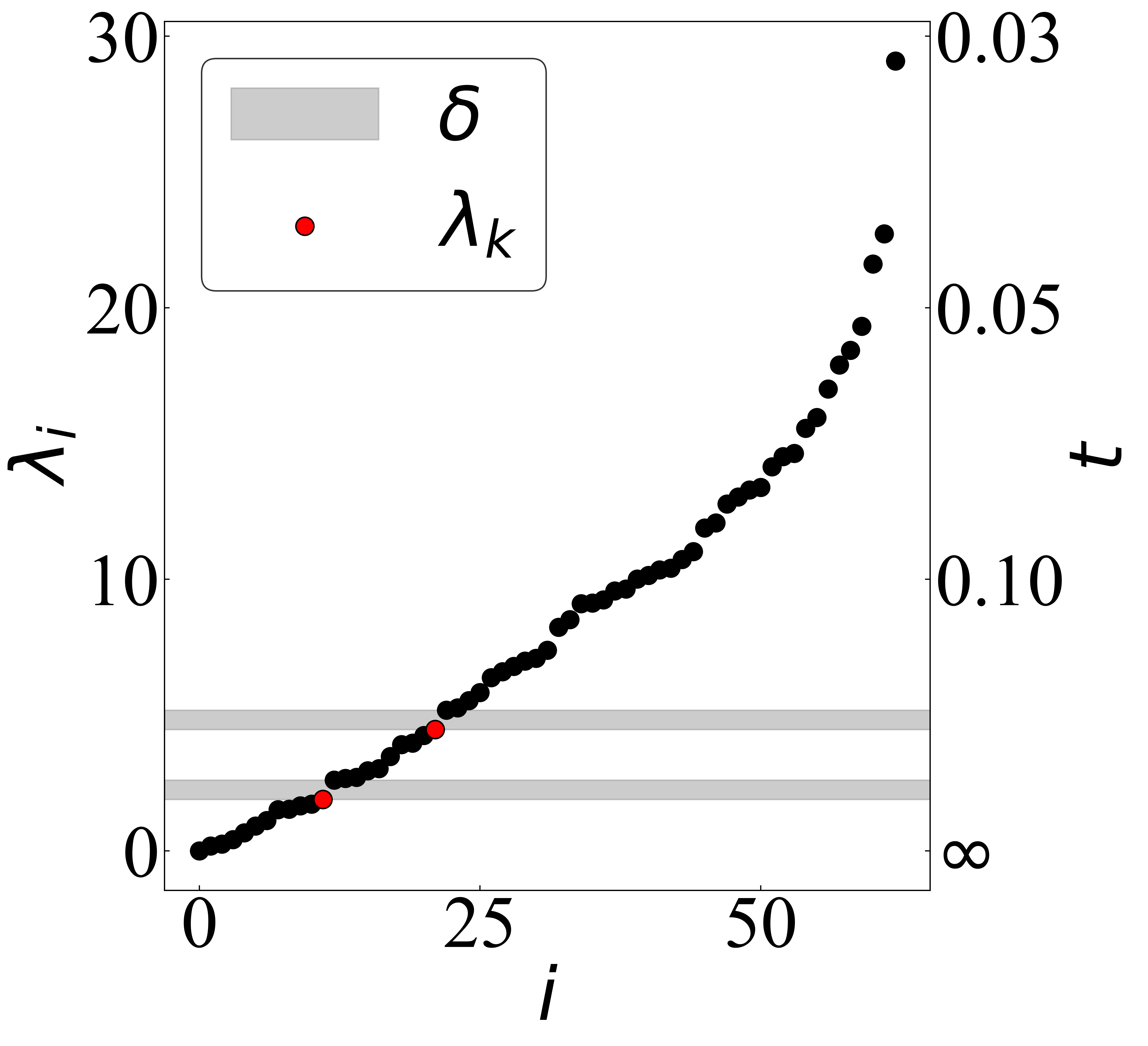}
    \end{tabular}
    \centering
    \includegraphics[width=0.9\linewidth]{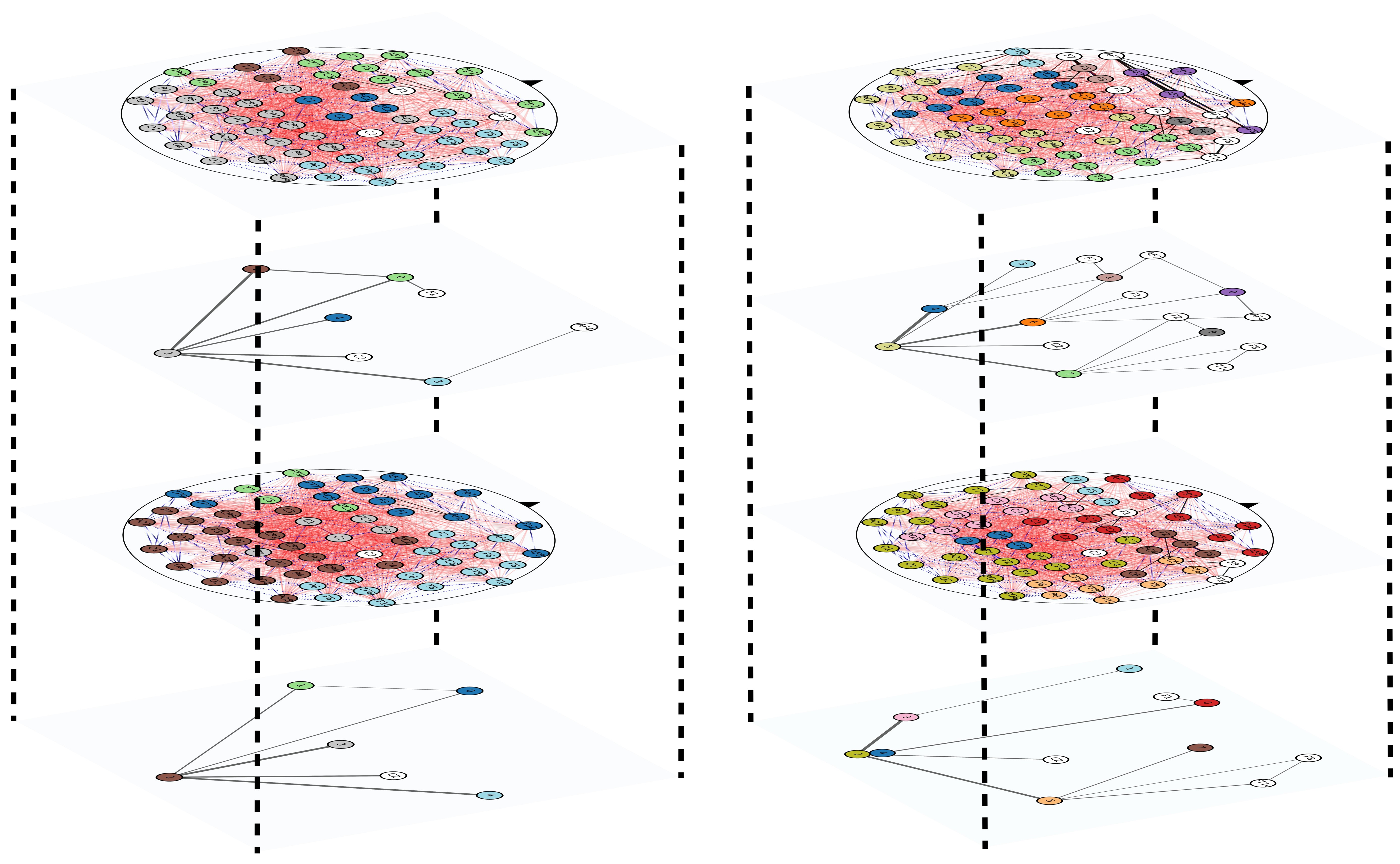} 
    \captionsetup{ font=small}
    \caption{\justifying Spectral gap and renormalization of the brain. For each $\lambda_{k}$, a renormalized graph is obtained. The electrode map shows the clusters of electrodes that are contracted into effective vertices, with colors showing which cluster belongs to which effective vertex. The effective vertices representing the occipital lobe are the most central in the renormalized graphs.
    }    
    \label{fig:Rest1andAttention1}
\end{figure*}

\textit{Introduction.}---
Graphs capture the intricate relationships and dynamics between the units that make up a complex system, such as the brain or financial markets. As such, they are powerful tools for analyzing and modeling these systems. However, different properties of complex systems emerge at different scales, making it useful to scale out of the graph structure using coarse-graining procedures. 
\\
\indent Coarse-graining scales out by aggregating clusters of vertices and edges, leading to a simplified representation that allows us to observe larger-scale structures while preserving key features of the original system. To ensure that these simplified representations remain true to the underlying dynamics of the system, we propose a diffusion-based coarse-graining and rescaling procedure for graphs, inspired by the renormalization group \cite{WILSON197475}, which does not rely on geometric closeness. This graph renormalization can reveal patterns and behaviors -- such as collective neural activity -- that may only become apparent at certain scales.
As we look at the graph Laplacian, our method follows the direction pointed out by \cite{Villegas_2023}'s Laplacian renormalization group scheme (LRG) and the spectral methods of \cite{Chung1997_Cheeger, Lambiotte_Schaub_2022, strogatz2000nonlinear}. However, instead of relying on the normalized heat kernel, we rely on the graph Laplacian to determine effective vertices and effective interactions. This results in a representation with both a coarse-grained structure and scale-adjusted weights.
Our method preserves the large-scale topological structure while better preserving the original diffusion dynamics. Since diffusion provides a notion of distance, this is consistent with the ideas of \cite{WILSON197475}, as we do not change the ``distance'' between the vertices or the ``size'' of the system in a physical sense.
\\ 
\indent The Renormalization Group (RG) \cite{PhysicsPhysiqueFizika.2.263,WILSON197475} allows us to study how collective behavior in complex systems arises at different scales. It is fundamental for understanding critical phenomena, phase transitions, universality, and scale dependence.
Intuitively, the RG is a tool to move from the microscopic to the macroscopic by coarse-graining and rescaling the system. At larger scales, the system's representation simplifies, as macroscopic behaviors are simpler and more universal than microscopic mechanisms \cite{PhysicsPhysiqueFizika.2.263, WILSON197475, RevModPhys.47.773, Wilson1979}. 
For systems, such as ferromagnets, coarse-graining relies on geometric distance in a metric space, as the strength of interactions tends to be highest between spins that are closest to each other. However, in many complex systems, closeness in a metric sense may not be the primary factor driving interactions. This is, for example, the case of the brain, composed of nonlinear units -- neurons -- interacting with each other and forming synaptic connections across distances. We consider it as a use case and construct a graph representation based on the mutual information between electrode signals from electroencephalogram (EEG) recordings, capturing the functional dependencies between different brain regions. This graph-based approach allows us to represent the interaction structure governing brain dynamics.
\\
\indent Applying the RG to such heterogeneous graphs poses significant challenges, as they do not have the homogeneous topology on which the RG relies. 
Important efforts besides \cite{Villegas_2023}'s LRG to extend the RG to graphs were made by \cite{PhysRevLett.95.171301}, which hypothesizes the existence of an embedding space responsible for the graph structure. This underlying space allows for identifying ``supernodes'', providing a coarse-grained graph description, akin to block spins in the real-space RG \cite{PhysicsPhysiqueFizika.2.263}. 
Also, the approach in \cite{García-Pérez2018} embeds graphs into underlying metric geometrical spaces. 
However, graphs are topological structures, thus a topological notion of the RG is needed \cite{Calcagni_2014}. Diffusion provides such a notion since a dynamical process can be defined on combinatorial structures and depends only on the structure’s topology. 
Furthermore, \cite{PhysRevResearch.4.033196} highlights that diffusion is linked to the graph scales through the graph communicability, with diffusion time serving as a resolution parameter \cite{MASUDA20171, Cimini2019}. The derivative of the entropy of the normalized heat kernel with respect to diffusion time indicates how fast a probability distribution becomes uniform. The largest entropy change over diffusion time corresponds to the graph's characteristic diffusion scale, which separates regions of fast diffusion from those where the diffusion slows down \cite{PhysRevResearch.4.033196}. This is the scale that the LRG zooms out to.  
We notice, from matrix perturbation theory and spectral clustering perspectives \cite{Stewart90_MatrixPertubation, vonluxburg2007tutorialspectralclustering, NIPS2001_801272ee},
that this scale is related to the spectral gap  $\delta = |\lambda_{k+1} - \lambda_{k}|$, defined as the largest difference between the consecutive smallest eigenvalues of the graph Laplacian \cite{Chung1997_Cheeger, Luxburg2007ATO}. The relation between the entropy and the spectral gap is further discussed in \cite{PhysRevX.6.041062}.
As $\lambda_{k}$ is inversely related to the diffusion time $t$, we determine the scale to zoom out to by heuristically identifying the spectral gap \cite{ NIPS2001_801272ee, Luxburg2007ATO}.
\\
\indent\textit{Renormalization.}--- To obtain a representation of the system for that characteristic scale, interactions must be adjusted to preserve both dynamics and important structural properties at larger scales. Thus, we introduce a renormalization procedure using a rescaled coarse-grained Laplacian that approximately preserves both. 
\\
\indent The rescaled coarse-grained Laplacian's ability to determine effective vertices and the effective weights between them can be understood by considering diffusion on an undirected graph $G = (V,E)$. The probability $\mathbf{p}(t)$ of the diffusion reaching one of the $|V|$ vertices at time $t$ is described by 
\begin{equation}\label{Mastereqaution}
    \frac{\partial}{\partial t}\mathbf{p}(t) = -\mathbf{L}\mathbf{p}(t), 
\end{equation}
where $\mathbf{L} = \mathbf{D} - \mathbf{A}\in \mathbb{R}^{|V| \times |V|}$ is the graph Laplacian matrix, with $\mathbf{A}$ being the weighted adjacency matrix and $\mathbf{D}$ a diagonal matrix with the strength of each vertex as entries. The variable $t$ is time rescaled by the diffusion constant and is therefore dimensionless. The entries of $\mathbf{p}(t)$ are the probabilities for the diffusion to reach one of the $|V|$ vertices at time $t$. 
Using the heat kernel, which is a matrix exponential $\mathrm{e}^{-t\mathbf{L}} = \sum_\alpha \frac{(-t\mathbf{L})^{\alpha}}{\alpha!}$, Eq.~(\ref{Mastereqaution}) is solved by 
\begin{equation}\label{SolutionMaster}
    \mathbf{p}(t) = \mathrm{e}^{-t\mathbf{L}}\mathbf{p}(0),
\end{equation}
where an entry $(\mathrm{e}^{-t\mathbf{L}})_{ij}$ considers all possible diffusion paths from vertex $i$ to $j$. It measures a diffusion-based closeness or similarity of vertices, giving more importance to larger-scale structures compared to the graph Laplacian. This is why it is used to identify clusters \cite{Chung1997_Cheeger, doi:10.1073/pnas.0708838104, Lambiotte_Schaub_2022}. 
\\
\indent Instead of using the heat kernel, we use a projection of the graph Laplacian into a subspace spanned by the eigenvectors corresponding to the smallest eigenvalues, which results in the coarse-grained Laplacian (see Eq.~\ref{NewLaplacian}). To understand why we can use the coarse-grained Laplacian, we project the graph Laplacian that is represented in $\mathcal{E} = \mathrm{span}(\{ \mathbf{e_l}\}_{l=0}^{|V|-1})$, where $\mathbf{e}_l\in \mathbb{R}^{|V|}$ is a standard basis vector, into space $\mathcal{S} = \mathrm{span}(\{ \mathbf{u}_k\}_{\alpha=0}^{|V|-1})$, where $\mathbf{u}_{\alpha}$ are the orthonormal eigenvectors of $\mathbf{L}$, via the spectral decomposition of the graph Laplacian $\mathbf{L} = \mathbf{U}\Lambda\mathbf{U}^{\top} = \sum_{\alpha}\lambda_{\alpha}\mathbf{u}_{\alpha}\mathbf{u}_{\alpha}^{\top}$. Since $\mathbf{L}$ is symmetric, the spectral theorem of symmetric matrices states that the eigenvectors are orthogonal to each other. After ensuring that each $\mathbf{u}_{\alpha}$ is normalized, $\mathbf{U}^{\top} = \mathbf{U}^{-1}$, where $\mathbf{U}$ has the eigenvectors $\mathbf{u}_\alpha$ as columns and the diagonal matrix $\Lambda$ has the corresponding eigenvalues $\lambda_\alpha$ on its diagonal. 
\\
\indent By considering the eigenvectors as modes that indicate graph partitions, we can see that the Laplacian in $\mathcal{S}$ provides insights into the graph structure at different scales:  
For a graph without disconnected components, the eigenvector $\mathbf{u}_0$ represents a mode of the graph's overall connectivity structure. The eigenvector $\mathbf{u}_1$, also known as the Fidler vector, represents a mode partitioning the graph into two. Higher-order eigenvectors $\mathbf{u}_{\alpha>1}$ correspond to modes representing, finer partitions of the graph, associated with sign and magnitude pattern in the mode $\mathbf{u}_{k>1}$ \cite{Chung1997_Cheeger}. So, smaller eigenvalues correspond to smoother modes and coarser partitions. 
\\
\indent This is explicitly shown by expressing $\mathbf{p}(t)$ as a superposition of exponentially varying modes using Eq.~(\ref{SolutionMaster}), $\mathbf{p}(t) = \sum_{\alpha}\mathrm{e}^{-t\lambda_\alpha}\,c_{\alpha}\mathbf{u}_\alpha$, where the initial state is represented as a linear combination of the eigenvectors $\mathbf{p}(0) = \sum_{\alpha}c_{\alpha}\mathbf{u}_\alpha$, and the coefficient $c_{\alpha}$ reflects how much a state aligns with $\mathbf{u}_{\alpha}$. Since $\mathbf{L}$ is hermitian, the eigenvalues are non-negative, and the modes are either stationary or exponentially decaying with time. As each eigenvector represents a diffusion mode for the evolution of the probability distribution across the vertices, modes associated with smaller eigenvalues $\lambda_{\alpha}$ decay slower and correspond to smoother eigenvectors. 
This allows us to zoom out and look at the behavior of the system over longer time scales by defining a coarse-graining procedure of the graph Laplacian using time-scale separation \cite{Chung1997_Cheeger, strogatz2000nonlinear, Lambiotte_Schaub_2022}. In essence, if the $k$-smallest eigenvalues are well separated from the remaining eigenvalues such that we have a large spectral gap $\delta = |\lambda_{k+1}-\lambda_{k}|$, we can simplify the description of the system. This is because the eigenmodes associated with eigenvalues larger than $\lambda_{k}$  become negligible for time-scales $t > 1/\lambda_{k+1}$, and we can describe the system by the $k$-slowest eigenmodes. They form a dominant subspace of the dynamics, which allows us to define a low dimensional approximate description of the dynamics of the graph \cite{Lambiotte_Schaub_2022}. Thus, we only consider the contributions of the k-slowest eigenmodes, resulting in the coarse-grained Laplacian 
\begin{equation}\label{NewLaplacian}
\mathbf{L}_{(1)} = \mathbf{U}_{(1)}\mathbf{\Lambda}_{(1)}\mathbf{U}^{\top}_{(1)} = \sum_{\alpha:\lambda_\alpha \leq \lambda_{k}} \lambda_\alpha\mathbf{u}_\alpha\mathbf{u}_{\alpha}^{\top}, 
\end{equation}
where $\mathbf{\Lambda}_{(1)}$ is a diagonal matrix of eigenvalues smaller than or equal to $\lambda_{k}$, and $\mathbf{U}_{(1)}$ is a matrix of the eigenvectors corresponding to these remaining eigenvalues, which form $\mathbf{L}_{(1)}$'s new basis $\mathcal{S}_{(1)} = \mathrm{span}(\{u_\beta\}_{\beta:\lambda_\beta \leq \lambda_{k}})$. The entries of $\mathbf{L}_{(1)}$ indicate how similar two vertices in $G$ are in the context of the remaining modes or from a zoomed-out perspective. This is because $(\mathbf{u}_{\alpha}\mathbf{u}_{\alpha}^{\top})_{ij} = u_{\alpha i}u_{\alpha j}$ is the product of the eigenvector components associated with vertices $i$ and $j$, which is the cosine similarity measure between these vertices within the mode $u_{\alpha}$. 
Thus, the sign and magnitude of the components $u_{\alpha i}$ and $u_{\alpha j}$ directly translate through $u_{\alpha i}u_{\alpha j}$ to how similar vertices $i$ and $j$ are within the structural pattern captured by $\mathbf{u}_{\alpha}$. 
Consequently, $(L_{(1)})_{ij}= \sum_{\alpha:\lambda_\alpha \leq \lambda_{k}}\lambda_\alpha u_{\alpha i}u_{\alpha j}$ is a ``weighted average'' of similarities between vertices $i$ and $j$, with eigenvalues $\lambda_{\alpha}$ acting as the ``weights''. 
\\
\indent Using Eq.~\eqref{NewLaplacian} and the definition of the graph Laplacian, we define 
\begin{equation}
    \mathbf{A}_{(1)} = \mathrm{diag}(\mathbf{L}_{(1)}) - \mathbf{L}_{(1)},
\end{equation} 
where $\mathrm{diag}(\mathbf{L}_{(1)})$ is a diagonal matrix containing the diagonal of $\mathbf{L}_{(1)}$. A positive entry in $\mathbf{L}_{(1)}$ or a negative entry in $\mathbf{A}_{(1)}$ indicates that vertices are similar in $G$ at a zoomed-out perspective and can be contracted. To illustrate this, we consider graph (a) in Fig. \ref{SynNetworks}, which consists of three 4-cliques with weights of 0.9, connected by edges with a weight of 0.1. The edges of $G_{(1)}$ with negative weights are highlighted in blue in Fig. \ref{SynNetworks}. 
To perform the contraction, we assign the respective similarity values from $\mathbf{A}_{(1)}$ as new weights to $\mathbf{A}$, which results in $\mathbf{A}_{(2)}$ with 
\begin{equation}\label{A2}
(\mathbf{A}_{(2)})_{ij} = \begin{cases}
(\mathbf{A}_{(1)})_{ij} & \text{if } (\mathbf{A})_{ij} > 0, \\
0 & \text{otherwise}.
\end{cases}
\end{equation}
Thereby, we only consider the weights of $\mathbf{A}_{(1)}$ from connections that also existed in $\mathbf{A}$. Then, we consider the corresponding graph $G_{(2)} = (V_{(2)},E_{(2)})$ (see Fig. \ref{SynNetworks}), and we contract the vertices $i \in V_{(2)}$ and $j \in V_{(2)}$ connected by a negative edge weight into effective vertices $s$, where $s$ remains connected to the vertices to which $i$ and $j$ were connected to. This results in the graph $G_{(3)} = (V_{(3)}, E_{(3)})$, which has the vertex set $V_{(3)} = (V_{(2)}\setminus\{i,j\})\,\bigcup\, s$ and $E_{(3)} = (E_{(2)}\setminus\{\mathrm{Edges}\,\mathrm{incident}\,\mathrm{to}\,i\,\mathrm{or}\,j\})\,\bigcup\, \{ (x,s_m \in s): (x,i)\in E_{(2)}\,\lor\,(x,j)\in E_{(2)}\}$. This preserves the structure of $G$ at a zoomed-out perspective (see Fig. \ref{SynNetworks}). 
Given that the graph Laplacian is an intensity matrix, it obeys the Perron-Frobenius theorem and therefore has at least one eigenvector with eigenvalue zero. For our renormalization procedure, the graph must be ergodic; otherwise, it may have oscillating modes that do not decay with time, and we do not have meaningful time-scale separation. 
Intuitively, these modes represent a random walker that is trapped and cannot explore the entire graph. Thus, the corresponding modes cannot represent meaningful graph partitions.
An ergodic graph has a non-degenerate zero eigenvalue. Consequently, for $\lambda_{k} \rightarrow 0$,  Eq.~\eqref{NewLaplacian} indicates that we will remain with the eigenvector associated with the eigenvalue zero, which tells us that all vertices are similar, and thus end up with a graph with one effective vertex. Conversely, for $\lambda_{k} \rightarrow \infty$, Eq.~\eqref{NewLaplacian} indicates that no eigenvectors are removed, and we remain with the original graph.
Intuitively, this coarse-graining procedure zooms out and decreases the resolution. Thus, we rescale the coarse-grained adjacency matrix $\mathbf{A}_{(3)}$ associated with $G_{(3)}$
\begin{equation}\label{FinalKSRG_eq}
    \mathbf{A}_{R} = \frac{1}{\lambda_{k}}\mathbf{A}_{(3)},
\end{equation}
which restores the original resolution. This means that the diffusion dynamics of the original system are approximately preserved in the renormalized system $G_{R}$. Since diffusion provides a notion of distance, this is analogous to the RG, where coarse-graining decreases the system size and increases the lattice spacing, while rescaling restores the original dimensions. From another perspective, rescaling ensures that the relative size of fluctuations in the new system matches those of the original system. We demonstrate this in Fig. \ref{DiffDySynNetworks} using Eq.~\eqref{SolutionMaster}. The approximation error satisfies 
\begin{equation}\label{Error}
\begin{split}
    \epsilon(t) &= \norm{\sum_{\alpha}\mathrm{e}^{-t\lambda_\alpha}\mathbf{u}_\alpha^{\top}\mathbf{p}(0)\mathbf{u}_\alpha - \sum_{\alpha:\lambda_\alpha \leq \lambda_{k}}\mathrm{e}^{-t\lambda_\alpha} \mathbf{u}_\alpha^{\top}\mathbf{p}(0)\mathbf{u}_\alpha} \\
    &\leq \mathrm{e}^{-t\lambda_{k+1}}\epsilon(0) 
    ,
\end{split}
\end{equation} 
where $\epsilon(0)$ is the initial error due to neglecting the fast modes. This error will decay exponentially with a rate of at least $\lambda_{k+1}$. Thus, the larger $\lambda_{k+1}$, the faster the decay of the $k+1$ modes, and the better the first $k$ modes describe the system dynamics. 
Furthermore, a large spectral gap $\delta$ leads to a faster decay of the fast modes relative to the slow ones, resulting in a clearer separation of times scales. Conversely, a small $\delta$ causes fast and slow modes to decay at similar rates. So, both contribute to the system dynamics for approximately the same duration, reducing the quality of the approximation. Thus, the quality of the approximation depends on both the decay rate $\lambda_{k+1}$ of the neglected modes and the spectral gap $\delta$, each influencing the error $\epsilon(t)$ in complementary ways \cite{Lambiotte_Schaub_2022, Chung1997_Cheeger, strogatz2000nonlinear}.
\\
\indent The Fig.~\ref{DiffDySynNetworks} compares the diffusion dynamics of the original graph (a) and the corresponding renormalized graph. In the original graph $G$, vertices contracted into effective vertices are highlighted by the colored circles (Fig. \ref{SynNetworks}). With Eq.~\eqref{SolutionMaster}, the diffusion probabilities $\mathbf{p}(t)$ are determined, where 
$\mathbf{p}(0)$ is a vector with all entries set to zero except for a value of one at one of the vertices within the red-highlighted cluster. This process is repeated for all vertices within the red cluster, and the average is shown in Fig.~\ref{DiffDySynNetworks}. These values are compared with the diffusion probabilities of the renormalized graph $G_{R}$, where $\mathbf{p}(0)$ is a vector with all entries set to zero except for a value of one at the effective vertex highlighted in red (Fig. \ref{SynNetworks}). Additionally, we compare to the graph $G^{LRG}$ obtained from \cite{Villegas_2023}'s Wilsonian LRG formulation, with the cut-off determined by the peak of the specific heat \cite{Villegas_2023, PhysRevResearch.4.033196}. Both our method and the Wilsonian LRG preserve the diffusion dynamics well in this example. However, the differences become more significant for more complex graphs, as we will see in example (b).  
If multiple characteristic scales $\lambda_{k}$ are present in the system, then by obtaining a renormalized graph for each, we move from the microscopic to a more macroscopic scale. This is shown in Fig.~\ref{BA_Net} for a Barabási–Albert graph (b) with 24 vertices, where the thick edges have weights of 0.9, and the thin edges weights of 0.1. The spectral gaps in Fig.~\ref{BA_Net} indicate two characteristic scales. Each renormalized graph in Fig.~\ref{BA_Net} preserves the zoomed-out structure of the graph, i.e., the skeleton. For example, for the first spectral gap $\lambda_{k}= 0.119$, seven effective vertices representing clusters in the original graph are formed: $s_{0}= \{0,5,8,11,15,16\}$, $s_{1} = \{1,20,14\}$, $s_{2} = \{2,6,22\}$, $s_{3} = \{3,4\}$, $s_{4} = \{7,13,17,19\}$, $s_{5} = \{9,10,12\}$, and $s_{6} = \{18,21,23\}$. These effective vertices provide insight into each scale's characteristic structures. 
\\
In comparison, the Louvain community detection algorithm finds similar communities, except it detects $\{2, 6, 9, 10, 12, 22\}$ as one community. This is not surprising, as it uses modularity optimization, which has a resolution limit. Thus, the Louvain algorithm may fail to detect small communities by merging them into larger clusters \cite{Fortunato_2007}. Additionally, not only the zoomed-out structure but also the diffusion dynamics of the original graph are preserved up to an error $\epsilon(t)$ (see Fig.~ \ref{DiffDySynNetworks}).
\textit{Human Neural Network Renormalization.}--- 
It is hypothesized that the brain self-organizes toward a critical state, exhibiting emergent properties and collective behaviors in neural activity across scales, such as neural avalanches \cite{Mora2011, PhysRevLett.59.381, PhysRevLett.108.208102, Beggs11167, PhysRevLett.125.028101}. To investigate this, we analyze brain activity across scales by renormalizing Triangulated Maximally Filtered Graphs (TMFG) \cite{10.1093/comnet/cnw015} extracted from the mutual information between electrode signals in \cite{GIFFORD2022119754}'s EEG recordings. We choose the TMFG as it is ergodic by definition. Each vertex represents the activity of a neuron group, and edge weights are the mutual information between the vertices they connect. 
The recordings were from a study where participants identified target images in a rapid sequence. They are divided into seven partitions: Rest 1 and Rest 2 (before and after the task) and Attention 1–5 (during the task). As expected, the spectrum reveals multiple spectral gaps, indicating multiple characteristic scales \cite{Bullmore2009ComplexBN}. We obtain a renormalized graph for the two smallest $\lambda_{k}$. This is visualized in Fig.~\ref{fig:Rest1andAttention1} for Rest 1 and Attention 1, which are representative of the brain’s rest and attention states. 
In the renormalized graphs, we observe the formation of effective vertices that absorb clusters of vertices where information spreads fast, indicating groups of strongly interacting neurons. These clusters highlight regions with high mutual information, suggesting functional neuronal connectivity and coordinated neuronal activity. 
\begin{figure}[H]
        \centering
        \begin{tabular}{c}
             Rest\\ 
            \includegraphics[width=0.6\linewidth]{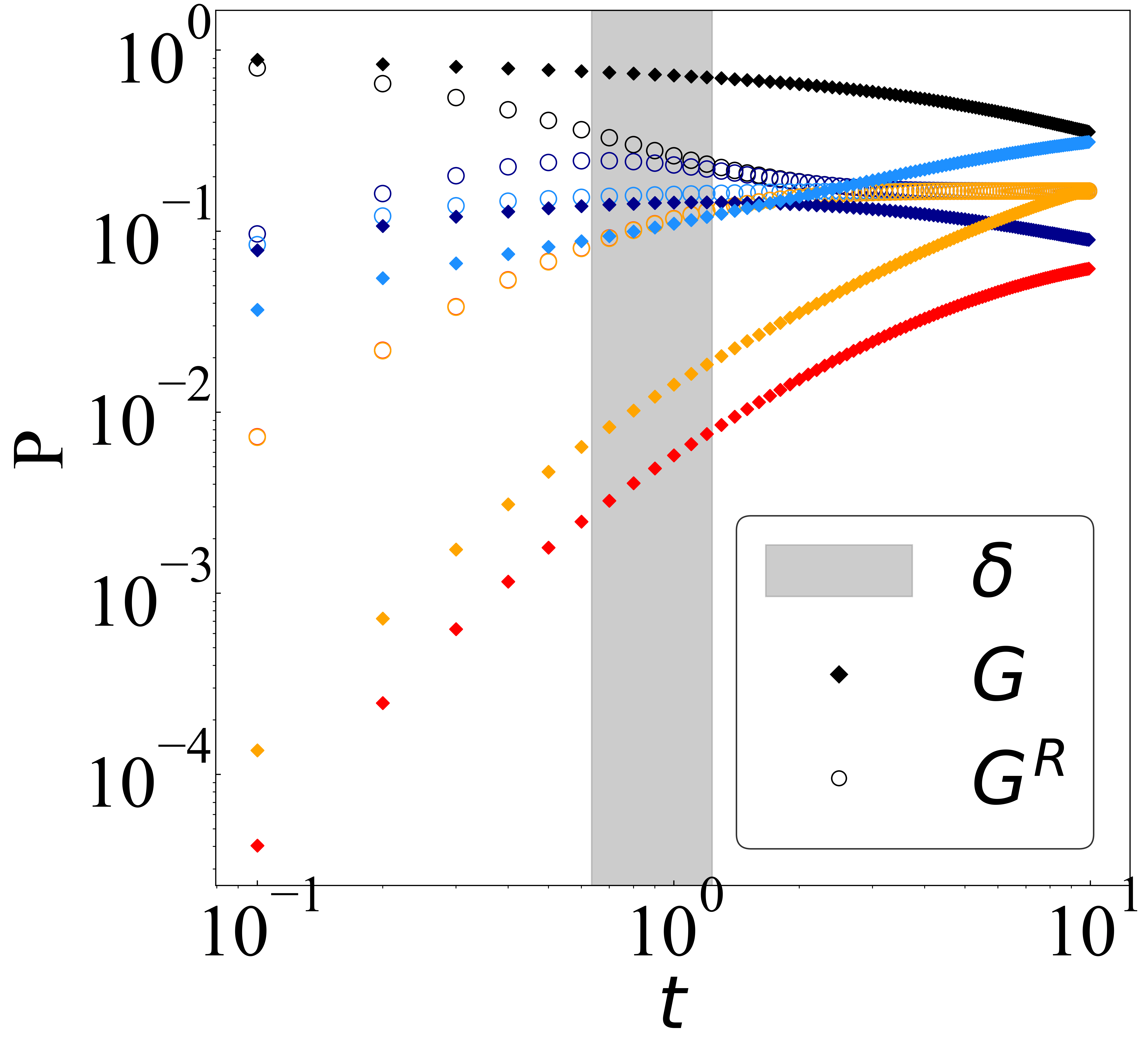}
            \\
             Attention \\
            \includegraphics[width=0.6\linewidth]{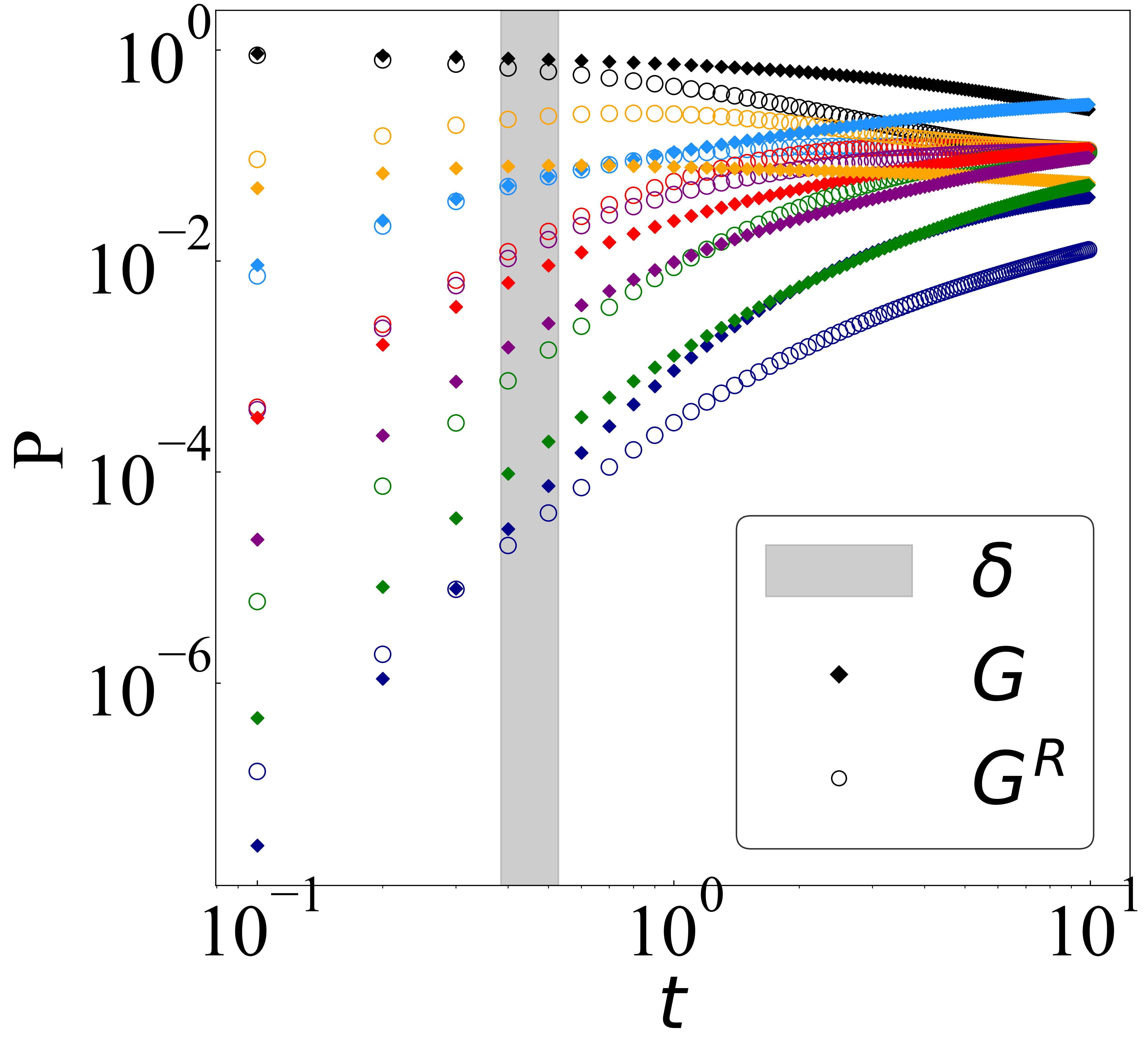}
        \end{tabular}
    \captionsetup{justification=justified, singlelinecheck=false, font=small}
    \caption{Diffusion dynamics of the brain. Transition probabilities between effective vertices in the renormalized graph $G^{R}$ and cluster in the original graph $G$ are compared using Eq.~\ref{SolutionMaster}. Diffusion starts from the effective vertex or cluster representing the frontal lobe. Each color indicates a different effective vertex or cluster.
    }
    \label{DiffDyBrainNetworks}
\end{figure}
Contracting these clusters into effective vertices reduces the redundant information shared by the vertices, resulting in a simplified representation of brain activity with significantly fewer vertices. Both resting states exhibit the same number of effective vertices absorbing a few large clusters, while the attention states have more, suggesting less redundant activation patterns, more flexible and efficient information processing, and specialization in brain dynamics \cite{Shew2010InformationCA}. 
This is supported by the clusters shape. In rest, the cluster covering most electrodes corresponding to the occipital lobe \{O1, O2, Oz, PO3, PO4, PO7, PO8\}, spans a wide, circular area in the posterior part of the electrode map, including parts of the parietal lobe, indicating less specialized and more generalized brain activity. 
In contrast, the corresponding cluster during attention, while still covering the occipital region, is more compact and irregular in shape, concentrated with fewer electrodes. These clusters are more specialized and localized, representing a more specific functional grouping and indicating a more focused and specialized processing of information. 
We observe that the diffusion dynamics are better preserved for attention states compared to rest states (see Fig. \ref{DiffDyBrainNetworks}). This is explained by $\lambda_{k+1}$ being smaller for the rest states than for the attention states (see Eq.~\ref{Error}). This suggests that, in rest states, macroscopic neural activation patterns do not align well with microscopic patterns, whereas for attention states, slow diffusion modes describe the pattern well across scales, indicating a higher degree of scale invariance.
\\
\indent\textit{Conclusions.}---
We have introduced a graph renormalization procedure based on the coarse-grained Laplacian. For different characteristic scales, identified by the spectral gap, it generates representations with reduced complexity (fewer vertices), reducing redundant information in graphs while preserving both their diffusion probabilities and the zoomed-out structure. Since most graph algorithms run in polynomial time relative to the number of vertices, this method facilitates the analysis of large graphs. 
\\
\indent It has allowed us to study the macroscopic properties of human brain activity emerging from the interactions of the system’s microscopic constituents. We have observed collective behavior in the form of clusters of coordinated neuronal activity. Additionally, the renormalized graphs show that brain activity dynamically reorganizes across scales during the object recognition task, with more generalized activity during rest and more specialized activity in the occipital lobe during attention. Moreover, our results suggest that activation states exhibit a higher degree of scale invariance, indicating that a simpler, and more universal representation of neural dynamics can be found via renormalization.  
\\
\indent\textit{Acknowledgments.}--- M. Schmidt thanks Renaud Lambiotte for his significant contributions and insightful comments on the manuscript, and Sabrina Aufiero, Peter Zhang, and Honyu Lin for their helpful discussions.
\bibliographystyle{apsrev4-1} 
\bibliography{bibliography}
\end{document}